\documentclass[12pt]{iopart}
\usepackage{iopams}  
\usepackage[T1]{fontenc}
\usepackage[latin9]{inputenc}
\usepackage{calc}
\usepackage{mathrsfs}
\usepackage{graphicx}
\usepackage{esint}

\makeatletter

\begin{document}
\title[SPOPO time/frequency quantum analysis]{A time/frequency quantum analysis of the light \\generated by synchronously pumped optical parametric oscillators}

\author{Shifeng Jiang, Nicolas Treps$^*$, and Claude Fabre}

\address{
Laboratoire Kastler Brossel, Université Pierre et Marie Curie - Paris 6, ENS, CNRS\\4 place Jussieu, 75252 Paris Cedex 05, France \\
$^*$Corresponding author: nicolas.treps@upmc.fr
}

\begin{abstract}
We present in this paper a general model to determine the quantum properties of the light generated by a synchronously pumped Optical Parametric Oscillator (SPOPO) operating below threshold. This model considers time and frequency on an equal footing, which allows us to find new quantum properties, related for example to the Carrier Envelope Offset phase (CEO), and to consider situations which are close to real experiments. We show that, in addition to multimode squeezing in the so-called ``supermodes'', the system exhibits quadrature entanglement between frequency combs of opposite CEO phases. We have also determined the quantum properties of the individual pulses, and their quantum correlations with the neighboring pulses. Finally we determine the Quantum-Cramer Rao limit for an ultra-short time delay measurement using a given number of pulses generated by the SPOPO.
 \end{abstract}



\noindent \section{Introduction}

\noindent  $\chi^{(2)}$ nonlinear media are efficient generators of non-classical states of light, such as squeezed states or quadrature entangled states, which are valuable resources for quantum metrology
and quantum information processing in the continuous-variable (C.V.)
regime \cite{Caves-81, Bachor-book}. These non-classical states of light are produced by  parametric down conversion either in the single pass regime \cite{Slusher-87} or inside a resonant cavity \cite{Wu-86}. In the latter case,  bright light is generated above some pump threshold value: one has an Optical Parametric Oscillator (OPO). Due to the resonant cavity buildup, OPOs
produce non-classical states of light with the moderate pump power delivered by c.w. lasers.  The quantum properties of c.w. OPOs have been extensively investigated over the past years, both theoretically and experimentally. In the non-degenerate case, they produce below and above threshold EPR entangled states of light \cite{Laurat-04,Ou-92,Coelho-09}. In the degenerate case they produce below threshold highly squeezed light in the resonant frequency and spatial modes which are resonant with the cavity \cite{Walls, Schnabel}. 

Recently there has been a rising interest for OPOs pumped by trains of ultrashort pulses, because of their promising properties: firstly, with such pumps, the efficiency of the nonlinear process can be
increased further and the oscillation threshold lowered; secondly the fact that the system generates light which spans over thousands of equally spaced frequency modes (``frequency combs``) opens the way to the efficient generation of highly multimode non-classical light \cite{Law-00, Wasilewski-06, Valcarcel-06}; in addition such sources can be used for the quantum metrology of ultra-short time delays \cite{Lamine-08}.

 Synchronously pumped OPO (SPOPO) are OPOs for which  the cavity round trip time
 is synchronized with the time interval between the pump pulses in order to favor the
resonant cavity-buildup. Such SPOPOs have been essentially developed as efficient classical sources of tunable ultrashort pulses \cite{Piskarskas-88,Edelstein-89,Cheung-90,McCarthy-93,Leindecker-11} in the femtosecond regime. They have also been used to generate squeezed light in the picosecond regime \cite{Shelby-92}, and more recently to produce multimode squeezed light in the femtosecond regime \cite{Pinel-11}.

Previous theoretical works on the quantum properties of SPOPOs
below threshold \cite{Valcarcel-06, Patera-10, Averchenko-11} have shown that SPOPOs generate
multimode squeezed frequency combs and that there exist quantum correlations
between pulses in different time intervals. These investigations have been carried out
either totally in the frequency domain (longitudinal modes), or totally in the time
domain. In this paper, we use the quantum input-output
approach and develop a formalism which allows us to treat the quantum properties of the SPOPO in the frequency and in the time domain on an equal footing. This enables us to find new interesting properties, related in particular to the Carrier Envelope Offset phase (CEO), and to present a comprehensive model, capable
of taking into account more realistic conditions, such as the intra-cavity
dispersion \cite{Jones-00}.

The paper is organized as follows. In Sect. 2, we first present an input-output formalism for treating OPOs in the general case. Then, in Sect. 3, we treat the case of an OPO pumped by a train of synchronous pulses, and derive the quantum properties of the field that is generated below threshold. In Sect. 4, the quantum properties of individual pulses are derived. Finally we evaluate in Sect. 5 the quantum improvement of the estimation of ultra-short time delays using the pulses generated by the SPOPO. Appendices give more technical details on the derivation of some of the presented results.

\noindent \section{Input-output relation for lossless degenerate ring-cavity OPOs below threshold}

\noindent Let us consider an OPO in a ring-cavity configuration, where
the cavity is transparent for the pump, and a $\chi^{(2)}$ crystal in
degenerate type-I configuration is placed in the cavity (see Fig.
1),
\begin{figure}[htb]
 \centerline{ \includegraphics[width=7cm]{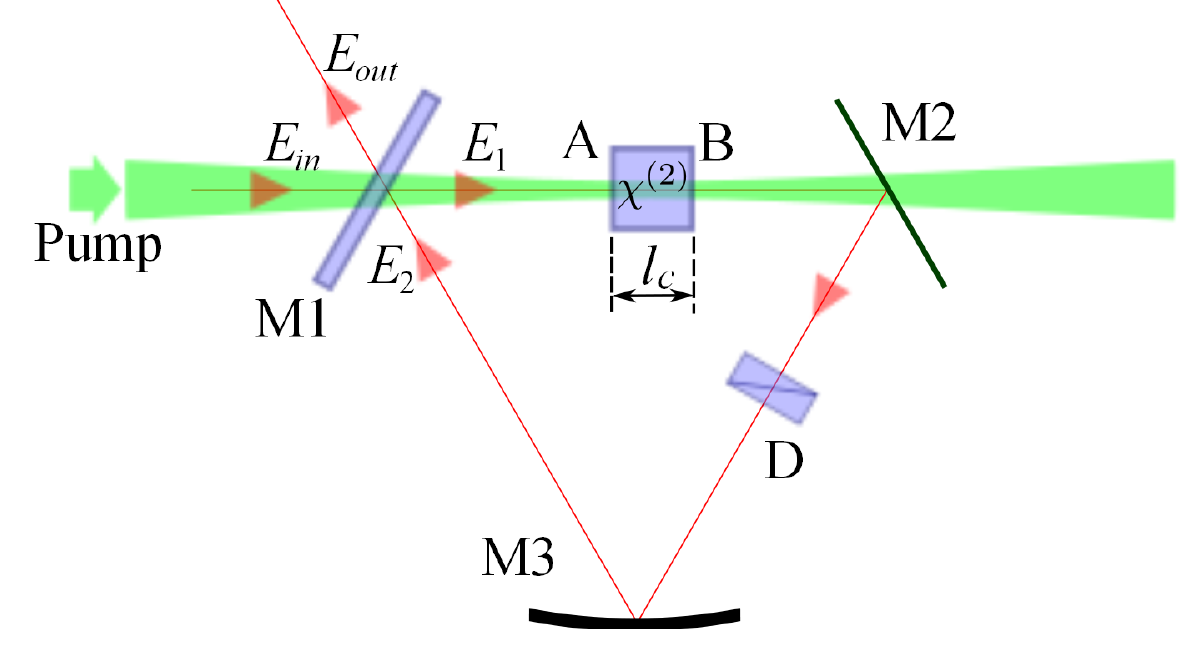}} \caption{Sketch of the SPOPO ring cavity}
\end{figure}
M1 is the output coupler, and D represents the group
of linear passive devices inserted for dispersion compensation. We will ignore
the transverse spatial dependence for the sake of simplicity. In the
free space outside the cavity, the positive-frequency components of
the incoming and outgoing quantum fields can be written, within the slowly
varying envelope approximation, as \cite{Kolobov-99} :
\begin{equation}\label{eq1}
\hat{E}_{\sigma}^{(+)}(t)=i\mathscr{E}_{0}\hat{a}_{\sigma}(t)e^{-i\omega_{0}t},\textrm{ with }\sigma\!=\! in,out,
\end{equation}
where $\mathscr{E}_{0}$ is the single photon electric field, $\omega{}_{0}$ is the
carrier frequency of both fields, and $\hat{a}_{in}$ and $\hat{a}_{out}$
are the slowly varying annihilation operators that verify the
commutation relations:
\begin{equation}\label{eq2}
[\hat{a}_{\sigma}\left(t\right),\hat{a}_{\sigma}\left(t'\right)]\!=\!0\textrm{ and }[\hat{a}_{\sigma}\left(t\right),\hat{a}_{\sigma}^{\dagger}\left(t'\right)]\!=\!\delta(t\!-\! t').
\end{equation}
One can also write the positive electric field operator in terms of annihilation operators of frequency modes, which are the Fourier transforms of time dependent annihilation operators and defined as: $\underline{\hat{a}}(\omega)=\int_{\Re}\hat{a}(t)e^{-i\omega t}{\rm d}t$, so that:
\begin{equation}\label{eq3}
[\underline{\hat{a}}_{\sigma}\left(\omega \right),\underline{\hat{a}}_{\sigma}^{\dagger}\left(\omega'\right)]\!=\!2 \pi\delta(\omega - \omega').
\end{equation}

Also for the sake of simplifying the model we assume that the losses are negligible. Thus,
since the pump depletion is negligible below threshold, the OPO that
we are dealing with is a linear lossless system, where the outgoing
(output) field only explicitly depends on the incoming (input) field.
Hence, the most general relation for our system reads, in the time as well as in the frequency domain:
\begin{equation}\label{eq4}
\left(\begin{array}{c}
\hat{a}_{out}\\
\hat{a}_{out}^{\dagger}
\end{array}\right)=\boldsymbol{\mathcal{R}}\left(\begin{array}{c}
\hat{a}_{in}\\
\hat{a}_{in}^{\dagger}
\end{array}\right),\textrm{ with }\boldsymbol{\mathcal{R}}=\left(\begin{array}{cc}
\mathcal{C} & \mathcal{S}\\
\mathcal{S}^{*} & \mathcal{C}^{*}
\end{array}\right)
\end{equation}

\noindent where
$\mathcal{C}$ and $\mathcal{S}$ are integral transforms. For example,
$\mathcal{C}\hat{a}_{in}$ reads $\int_{\Re}C(t,t')\hat{a}_{in}(t'){\rm d}t'$
in the time domain and $\int_{\Re}\underline{C}(\omega,\omega')\underline{\hat{a}}_{in}(\omega'){\rm d}\omega'$
in the frequency domain, where $C$ is the integral kernel
\footnote{\noindent  In this paper, bold letters will denote matrix and vectors,
and calligraphic letters will denote integral transforms. The underline
will denote either the Fourier transforms, defined as: $\underline{\alpha}(\omega)=\int_{\Re}\alpha(t)e^{-i\omega t}{\rm d}t$,
or the kernels in the frequency domain.%
}. Since relation (\ref{eq4}) must preserve commutation relations (\ref{eq2}), the
above linear Bogoliubov transform should be complex symplectic (hereinafter, simply called symplectic), equivalent
to a $2\times2$ real symplectic transform \cite{Dutta-95}. The necessary
and sufficient condition for this is: 
\begin{equation}\label{eq5}
\boldsymbol{\mathcal{R}}^{-1}\!=\!\boldsymbol{\sigma}_{1}\boldsymbol{\mathcal{R}}^{\dagger}\boldsymbol{\sigma}_{1},\textrm{ with }\boldsymbol{\sigma}_{1}=\left(\begin{array}{cc}
1 & 0\\
0 & -1
\end{array}\right).
\end{equation}

\noindent Note that the effect of linear losses or other cavity
configurations can also be described by using a symplectic transform,
where the dimension of system is increased while the symplectic character of the transformation
is preserved. 

For our linear and lossless system, the expression of $\boldsymbol{\mathcal{R}}$
can be found with the help of the relations between classical fields \cite{Walls}.
We denote $\boldsymbol{\mathcal{T}}_{{\rm RT}}$ the cavity round-trip
transform, which relates the internal field envelopes after and before the
output coupler M1, i.e, 
\begin{equation}\label{eq6}
\left(\begin{array}{c}
A_{1}\\
A_{1}^{*}
\end{array}\right)=\boldsymbol{\mathcal{T}}_{{\rm RT}}\left(\begin{array}{c}
A_{2}\\
A_{2}^{*}
\end{array}\right),
\end{equation}
\noindent where $A_{1,2}(t)=E_{1,2}^{(+)}(t)\exp(i\omega_{0}t)$ are
the slowly varying envelopes. Since (\ref{eq6}) also holds without the output
coupler M1, $\boldsymbol{\mathcal{T}}_{{\rm RT}}$ is
symplectic. We can write: 
\begin{equation}\label{eq7}
\boldsymbol{\mathcal{T}}_{{\rm RT}}\!=\!\boldsymbol{\mathcal{U}}_{{\rm 1B}}\boldsymbol{\mathcal{S}}_{{\rm BA}}\boldsymbol{\mathcal{U}}_{{\rm A1}},
\end{equation}
where $\boldsymbol{\mathcal{S}}_{{\rm BA}}$ is the transform of the
$\chi^{(2)}-$crystal, and $\boldsymbol{\mathcal{U}}_{{\rm A1}}$
and $\boldsymbol{\mathcal{U}}_{{\rm 1B}}$ are transforms from M1
to point A, and from B to M1. These transforms must all be symplectic,
and their explicit expressions will be discussed later. At M1, the
boundary condition for lossless coupling of the internal and external
fields can be written as:
\begin{equation}\label{eq8}
\left(\begin{array}{c}
E_{1}\\
E_{out}
\end{array}\right)=\left(\begin{array}{cc}
t & r\\
-r & t
\end{array}\right)\left(\begin{array}{c}
E_{in}\\
E_{2}
\end{array}\right),
\end{equation}
where $t$ and $r$ are real amplitude transmission and reflection
coefficients ($r^{2}\!+\! t^{2}\!=\!1$), which are assumed to be
constant in the frequency range of interest. Then, with (\ref{eq6}) and (\ref{eq8}),
we find:
\begin{equation}\label{eq9}
\boldsymbol{\mathcal{R}}=\frac{\boldsymbol{\mathcal{T}}_{{\rm RT}}-r}{1-r\boldsymbol{\mathcal{T}}_{{\rm RT}}}=r^{-1}\left(\frac{t^{2}}{1-r\boldsymbol{\mathcal{T}}_{{\rm RT}}}-1\right),
\end{equation}
which, formally independent of the pumping condition, can be considered
as a generalization of previous works (see, for example, Eq. (5.13b)
of \cite{Reynaud}). From (\ref{eq9}), it is easy to verify that $\boldsymbol{\mathcal{R}}(\boldsymbol{\mathcal{T}}_{{\rm RT}})=\boldsymbol{\mathcal{R}}^{-1}(\boldsymbol{\mathcal{T}}_{{\rm RT}}^{-1})$.
Then, since $\boldsymbol{\mathcal{T}}_{{\rm RT}}$ is symplectic,
one can show with (\ref{eq5}) that $\boldsymbol{\mathcal{R}}$ is symplectic.

Outside the crystal, the transforms $\boldsymbol{\mathcal{U}}_{{\rm A1}}$
and $\boldsymbol{\mathcal{U}}_{{\rm 1B}}$, which are defined in (\ref{eq7}),
describe the lossless linear dispersive propagations. They are unitary
diagonal, for example: $\boldsymbol{\mathcal{U}}_{{\rm A1}}\!=\!{\rm Diag}(\mathcal{U}_{{\rm A1}},\mathcal{U}_{{\rm A1}}^{*})$,
where $\mathcal{U}_{{\rm A1}}$ is unitary and its kernel reads: 
\begin{equation}\label{eq10}
\underline{U}_{{\rm A1}}(\omega,\omega')\!=\!\delta(\omega-\omega')\exp[i\phi_{{\rm A1}}(\omega)],
\end{equation}
$\phi_{{\rm A1}}(\omega)$ being the spectral phase modification
due to the propagation from M1 to A.

In the lossless crystal, which
extends from $z_{{\rm A}}\!=\!-l_{c}/2$ to $z_{{\rm B}}\!=\! l_{c}/2$,
with the non-depleted pump approximation, the signal propagation can
be described by the linear evolution of its slowly varying envelope:
$(A_{s}(z),A_{s}^{*}(z))^{{\rm T}}=\boldsymbol{\mathcal{S}}(z)(A_{s}(z_{{\rm A}}),A_{s}^{*}(z_{{\rm A}}))^{{\rm T}}$.
So, the transform describing the propagation in the $\chi^{(2)}-$crystal is given by: $\boldsymbol{\mathcal{S}}_{{\rm BA}}\!=\!\boldsymbol{\mathcal{S}}(\boldsymbol{z}_{{\rm B}})$.
From the classical description of traveling-wave degenerate parametric down conversion (PDC) \cite{Wasilewski-06, Yariv-88}, $\boldsymbol{\mathcal{S}}(z)$ can be found by solving
\begin{equation}\label{eq11}
\partial_{z}\boldsymbol{\mathcal{S}}(z)=\boldsymbol{\mathcal{K}}(z)\boldsymbol{\mathcal{S}}(z),\textrm{ }\boldsymbol{\mathcal{S}}(z_{{\rm A}})=1,
\end{equation}
with the kernel of $\boldsymbol{\mathcal{K}}(z)$ given by:
\begin{equation}\label{eq12}
\mathbf{K}(z,t,t')\!=\!\left(\!\begin{array}{cc}
-ik_{s}(-i\partial_{t}) & 2\pi\chi_{0}A_{p}(z,t)\\
2\pi\chi_{0}A_{p}^{*}(z,t) & ik_{s}(i\partial_{t})
\end{array}\!\right)\!\delta(t-t'),
\end{equation}

\noindent In this relation the diagonal part describes the free propagation in
the lossless medium and $k_{s}\left(-i\partial_{t}\right)$ is the signal wavevector $k_{s}(\omega_{0}+ \omega)$ around $\omega_{0}$  in which $\omega$ has been replaced by
$-i\partial_{t}$, whereas the anti-diagonal part describes the PDC
process. Because $\boldsymbol{\sigma}_{1}\mathcal{\boldsymbol{K}}^{\dagger}(z)\boldsymbol{\sigma}_{1}\!=\!-\mathcal{\boldsymbol{K}}(z)$,
we have $\boldsymbol{\sigma}_{1}\boldsymbol{\mathcal{S}}^{\dagger}(z)\boldsymbol{\sigma}_{1}\boldsymbol{\mathcal{S}}(z)\equiv1$.
Thus, from (\ref{eq4}), wee see that (\ref{eq11}) and (\ref{eq12}) describe a symplectic evolution. 

The effective nonlinear
coefficient $\chi_{0}$ is given by:
\begin{equation}\label{eq13}
\chi_{0}=\left(\frac{2\omega_{0}^{2}}{\epsilon_{0}n_{0}^{3}c^{3}\mathscr{A}_{{\rm eff}}}\right)^{\frac{1}{2}}d_{{\rm eff}},
\end{equation}
where $n_{0}=ck_{s}(\omega_{0})/\omega_{0}$ is the mean refractive index
for the signal wave, $\mathscr{A}_{{\rm eff}}$ is the effective area of degenerate
three-wave mixing, and $d_{{\rm eff}}$ is the effective nonlinearity
coefficient for the given phase-matching configuration. The pump field
envelope reads in the frequency domain: $\underline{A}_{p}(z,\omega)\!=\!\underline{A}_{p}(\omega)\mathrm{exp}[-ik_{p}(\omega)z]$,
where $k{}_{p}\left(\omega\right)$ is the pump wave-number at optical
frequency $2\omega_{0}+\omega$. It is normalized so that $|A{}_{p}(z,t)|^{2}$
is the pump power.

For OPOs, as long as the cavity finesse is relatively high, the amount
of round-trip PDC is generally small below threshold. So, an approximate
symplectic solution of (\ref{eq11}) and (\ref{eq12}) can be adopted, which is in fact
the first order approximation of a Magnus expansion, also
known as exponential perturbation theory, that preserves the symplectic character of the transformation \cite{Blanes-09}: 
\begin{equation}\label{eq14}
\boldsymbol{\mathcal{S}}_{{\rm BA}}\!=\!\boldsymbol{\mathcal{U}}_{{\rm BA}}\boldsymbol{\mathcal{G}}_{0},
\end{equation}
where $\boldsymbol{\mathcal{U}}_{{\rm BA}}$ is crystal's unitary
transform without pumping, and
\begin{equation}\label{eq15}
\boldsymbol{\mathcal{G}}_{0}={\rm exp}\left(\begin{array}{cc}
0 & \mathcal{S}_{0}\\
\mathcal{S}_{0}^{\mathtt{*}} & 0
\end{array}\right),
\end{equation}
with
\begin{equation}\label{eq16}
\underline{S}_{0}\left(\omega,\omega'\right)=\chi_{0}l_{c}\underline{A}_{p}(\omega+\omega')\Phi(\omega,\omega'),
\end{equation}
where $\Phi(\omega,\omega')=\sin\Delta\phi(\omega,\omega')/\Delta\phi(\omega,\omega')$
is the phase-matching function with $\Delta\phi(\omega,\omega')\!=\! l_{c}[k_{s}(\omega)\!+\! k_{s}(\omega')\!-\! k_{p}(\omega\!+\!\omega')]/2$.
Since $\mathcal{S}_{0}$ is symmetric, $\boldsymbol{\mathcal{G}}_{{\rm 0}}$
is symplectic and positive-definite. Thus, from (\ref{eq7}) and (\ref{eq14}), we can
write: $\boldsymbol{\mathcal{T}}_{{\rm RT}}\!=\!\boldsymbol{\mathcal{U}}_{{\rm RT}}\boldsymbol{\mathcal{G}}_{{\rm RT}}$,
where $\boldsymbol{\mathcal{U}}_{{\rm RT}}\!=\!\boldsymbol{\mathcal{U}}_{{\rm 1B}}\boldsymbol{\mathcal{U}}_{{\rm BA}}\boldsymbol{\mathcal{U}}_{{\rm A1}}$
is the round-trip unitary transform without pumping, and $\boldsymbol{\mathcal{G}}_{{\rm RT}}\!=\!\boldsymbol{\mathcal{U}}_{{\rm A1}}^{\dagger}\boldsymbol{\mathcal{G}}_{0}\boldsymbol{\mathcal{U}}_{{\rm A1}}$
is the positive transform of round-trip parametric gain. In
the following we will
replace $\boldsymbol{\mathcal{G}}_{{\rm RT}}$ by $\boldsymbol{\mathcal{G}}_{0}$, which amounts to performing a unitary transform ($\hat{a}_{\sigma}\rightarrow\mathcal{U}_{{\rm A1}}\hat{a}_{\sigma}$) simultaneously
on the input and the output.
So, (\ref{eq9}) can be rewritten as:
\begin{equation}\label{eq17}
\boldsymbol{\mathcal{R}}=r^{-1}\left(\frac{t^{2}}{1-r\boldsymbol{\mathcal{U}}_{{\rm RT}}\boldsymbol{\mathcal{G}}_{{\rm 0}}}-1\right).
\end{equation}

The round-trip unitary transform $\boldsymbol{\mathcal{U}}_{{\rm RT}}$
describes the passive round-trip spectral phase modification, $\phi{}_{{\rm RT}}(\omega)$.
Around the signal carrier frequency $\omega{}_{0}$, it can be approximated
as a polynomial of $\omega$. For the sake of simplicity, we will
assume the higher order dispersions are compensated or negligible
so that 
\begin{equation}\label{eq18}
\phi{}_{{\rm RT}}(\omega)\!=\!\Delta_{{\rm RT}}\!+\!\omega T_{0},
\end{equation}
where $T_{0}$ is the round-trip time, and $\Delta{}_{{\rm RT}}$
is the round-trip phase shift that detunes the resonance of cavity.
Therefore, the kernel of $\boldsymbol{\mathcal{U}}_{{\rm RT}}$ reads in the time domain:
\begin{equation}\label{eq19}
\mathbf{U}_{{\rm RT}}(t,t')=\delta(t-t'+T_{0})e^{i\boldsymbol{\sigma}_{1}\Delta_{{\rm RT}}},
\end{equation}
which describes a time translation with a phase shift on the field
envelope.

\noindent \section{Frequency combs as a basis for analyzing SPOPOs quantum properties}

\noindent We will now apply the general formalism of Section 2 to the specific case of SPOPOs, in which the pump pulses are synchronized to the cavity, so that the pump field envelope can be written as:
\begin{equation}\label{eq20}
A_{p}(t)=\sqrt{\mathscr{E}_{p}}\sum_{k}e^{2ik\Delta_{0}}\alpha_{p}(t-kT_{{\rm 0}}),
\end{equation}
where $\mathscr{E}_{p}$ is the pump pulse energy, $2\Delta_{0}$
is the pump CEO, and $\alpha_{p}(t)$ is the pulse envelope function,
non-zero on $\mathscr{T}_{0}=[0,\textrm{ }T_{0}]$ and normalized, such
that $\int_{\mathscr{\mathscr{T}}_{0}}|\alpha_{p}(t)|^{2}{\rm d}t=1$.
Since the pump field is quasi $T_{0}-$periodic: $A_{p}(t+kT_{{\rm 0}})=A_{p}(t)e^{2ik\Delta_{0}},$
we have a quasi periodic system. From (\ref{eq11}) and (\ref{eq12}), we see that the
kernels of $\boldsymbol{\mathcal{S}}_{{\rm BA}}$, $\boldsymbol{\mathcal{G}}_{{\rm 0}}$
and $\mathcal{S}_{0}$ are quasi periodic translation invariant, e.g.,
\begin{equation}\label{eq21}
\mathbf{G}_{0}(t+T_{0},t'+T_{0})\!=\! e^{i\boldsymbol{\sigma}_{1}\Delta_{0}}\mathbf{G}_{{\rm 0}}(t,t')e^{-i\boldsymbol{\sigma}_{1}\Delta_{0}}.
\end{equation}
With (\ref{eq20}), we can rewrite (\ref{eq16}) in the time domain as:
\begin{equation}\label{eq22}
S_{0}(t,t')=\sum_{k}S_{c}(t-kT_{0},t'-kT_{0})e^{2ik\Delta_{0}},
\end{equation}
where the kernel of $\mathcal{S}_{c}$ reads in the frequency domain:
\begin{equation}\label{eq23}
\underline{S}_{c}\left(\omega,\omega'\right)=\chi_{0}l_{c}\mathscr{E}_{p}^{1/2}\underline{\alpha}_{p}(\omega+\omega')\Phi(\omega,\omega').
\end{equation}
In what follows, we will assume 
\begin{equation}\label{eq24}
S_{c}(t,t')=0,\textrm{ }\forall t,t'\notin\mathscr{T}_{0},
\end{equation}
the validity of which requires that the spectral width of the phase matching
function $\Phi(\omega,\omega')$ is much larger than the repetition
rate $T_{0}^{-1}$, and that the pump pulses are localized around the center
of each period and have a temporal width much smaller than $T_{0}$.
These conditions are generally satisfied in practice with SPOPOs ultrashort
pulses. Physically, (\ref{eq24}) implies that the signal photons down-converted
from the pump photons in each period will not travel into other periods
when leaving the crystal. 

Due to the quasi-periodicity property (\ref{eq21}), $\boldsymbol{\mathcal{G}}_{0}$
commutes with a unitary pseudo-time translation $\boldsymbol{\mathcal{U}}_{0}$
whose kernel reads:
\begin{eqnarray}\label{eq25}
\mathbf{U}_{0}\left(t,t'\right) & = &\delta\left(t-t'+T_{0}\right)e^{-i\boldsymbol{\sigma}_{1}\Delta_{0}}\nonumber \\
 & = & \mathbf{U}_{{\rm RT}}\left(t,t'\right)e^{-i\boldsymbol{\sigma}_{1}(\Delta_{{\rm RT}}+\Delta_{0})},
\end{eqnarray}
where the second line comes from (\ref{eq19}). So, $\boldsymbol{\mathcal{G}}_{0}$
and $\boldsymbol{\mathcal{U}}_{0}$ have common eigenvector functions.
The eigenvalue problem of these two transforms is discussed in the
Appendix A. With condition (\ref{eq24}), we have: 
\begin{eqnarray}\label{eq26}
\boldsymbol{\mathcal{G}}_{{\rm 0}}\boldsymbol{f}_{n}^{(\pm)}(\theta) & = & e^{\pm g_{n}}\boldsymbol{f}_{n}^{(\pm)}(\theta),\\
\boldsymbol{\mathcal{U}}_{{\rm 0}}\boldsymbol{f}_{n}^{(\pm)}(\theta) & = & e^{i\theta}\boldsymbol{f}_{n}^{(\pm)}(\theta),
\end{eqnarray}
with $g_{n}\geq0$, $\theta\in[-\pi,\pi]$ and 
\begin{equation}\label{eq28}
\boldsymbol{f}_{n}^{(\pm)}\left(\theta,t\right)=\frac{1}{\sqrt{2}}\left(\begin{array}{c}
f_{n}\left(\theta,t\right)\\
\pm f_{n}^{*}\left(-\theta,t\right)
\end{array}\right).
\end{equation}
In the above expression, $f_{n}(\theta,t)$ is the eigenfunction of
$\mathcal{S}_{0}\mathcal{S}_{0}^{\dagger}$ and reads:
\begin{equation}\label{eq29}
f_{n}(\theta,t)=\frac{1}{\sqrt{2\pi}}\sum\limits _{k=-\infty}^{+\infty}\psi_{n}(t-kT_{0})e^{ik(\theta+\Delta_{0})},
\end{equation}
with $\psi_{n}(t)\!=\!0$, $\forall t\notin\mathscr{T}_{0}$, where
$\psi_{n}(t)$ is the eigenfunction of $\mathcal{S}_{c}\mathcal{S}_{c}^{\dagger}$
with eigenvalue $g_{n}^{2}$. The set \{$\psi_{n}$\} is orthonormal
and complete over $\mathscr{T}_{0}$. 

We see that each function $f_{n}(\theta,t)$ represents a continuous
train of pulses, with a CEO of $\theta+\Delta_{0}$. In the frequency
domain, as shown on Fig. 2,
\begin{figure}[htb]
 \centerline{ \includegraphics[scale=0.7]{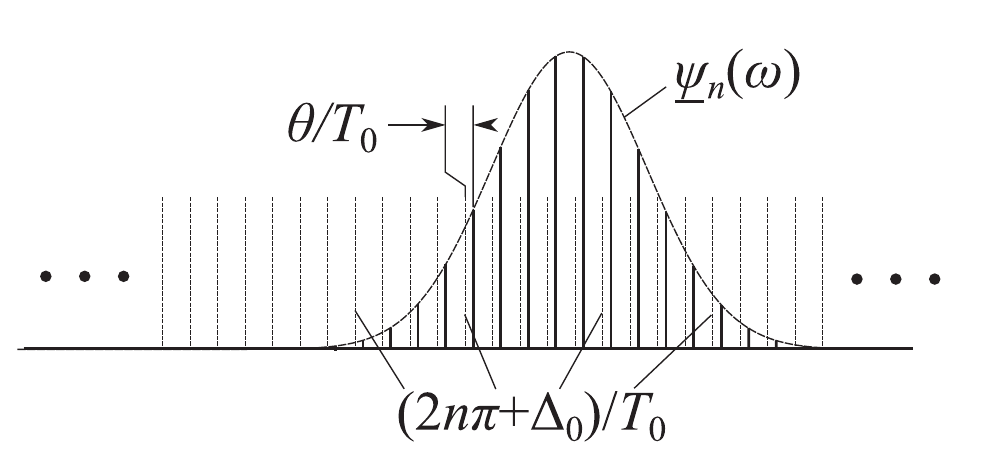}} \caption{Spectrum of a frequency shifted frequency comb.}
\end{figure}
 $\underline{f_{n}}(\theta,\omega)$ is a frequency comb shifted by
$\theta T_{0}^{-1}$ from the central frequencies of PDC $(2n\pi+\Delta_{0})/T_{0}$.
The envelope functions \{$\psi_{n}$\} are the so-called supermodes
\cite{Patera-10} and have also been studied in the context of single-pass
and single pump pulse \cite{Law-00, Wasilewski-06}, i.e., $T_{0}\rightarrow\infty$.
One contribution of this work is the introduction of the parameter
of frequency shift, $\theta$, that is proper for describing frequency
combs. In Appendix B, we show that \{$f_{n}(\theta,t)$\} forms
a complete orthonormal set over the real time axis. So, the slowly
varying envelope operators can be decomposed on the basis of frequency
combs: 
\begin{equation}\label{eq30}
\hat{a}(t)\!=\!\sum_{n}\int_{-\pi}^{+\pi}\hat{a}_{n}(\theta)f_{n}(\theta,t){\rm d}\theta,
\end{equation}
where $\hat{a}_{n}(\theta)=\int_{-\infty}^{+\infty}f_{n}^{*}(\theta,t)\hat{a}(t){\rm d}t$
is the frequency comb annihilation operator, whose commutation relations read: $[\hat{a}_{n}(\theta),\hat{a}_{m}(\theta')]\!=\!0$
and $[\hat{a}_{n}(\theta),\hat{a}_{m}^{\dagger}(\theta')]\!=\!\delta_{nm}\delta(\theta\!-\!\theta')$. 

Now, on the basis of frequency combs, with (\ref{eq25},\ref{eq26}), we are ready
to perform a Bloch-Messiah reduction \cite{Braunstein-05} on the transform (\ref{eq17}).
This yields:
\begin{equation}\label{eq31}
\left(\begin{array}{c}
\hat{a}_{out,n}(\theta)\\
\hat{a}_{out,n}^{\dagger}(-\theta)
\end{array}\right)=\frac{e^{i\theta}\mathbf{T}_{n}-r}{\mathbf{1}-re^{i\theta}\mathbf{T}_{n}}\left(\!\begin{array}{c}
\hat{a}_{in,n}(\theta)\\
\hat{a}_{in,n}^{\dagger}(-\theta)
\end{array}\right),
\end{equation}
with 
\begin{equation}\label{eq32}
\mathbf{T}_{n}=e^{i\boldsymbol{\sigma}_{1}(\Delta_{{\rm RT}}+\Delta_{0})}\left(\begin{array}{cc}
{\rm \cosh}g_{n} & {\rm \sinh}g_{n}\\
{\rm \sinh}g_{n} & {\rm \cosh}g_{n}
\end{array}\right).
\end{equation}
We can rewrite (\ref{eq31}) under the form: 
\begin{equation} \label{squeeze}
\hat{a}_{out,n}(\theta)=C_{n}(\theta)\hat{a}_{in,n}(\theta)+S_{n}(\theta)\hat{a}_{in,n}^{\dagger}(-\theta)
\end{equation}
with $|C_{n}(0)|^{2}-|S_{n}(0)|^{2}=1$. Thus the input-output relation (\ref{squeeze}) describes a squeezing transformation if $\theta=0$ and a twin photon generation when $\theta \neq 0$. We conclude that SPOPOs
are multimode squeezers for frequency combs with zero frequency shift
from the central frequencies of PDC. In the
time domain, these frequency combs are continuous trains of pulses
with a CEO half of that of the pump. For $\theta\neq0$, SPOPOs generate
pairs of entangled frequency combs with opposite frequency shifts,
or in the time domain continuous trains of pulses whose CEOs are symmetric with respect
to $\Delta_{0}$, i.e., $\Delta_{0}\pm\theta$. This is just as in
the c.w. case where the c.w. modes with optical frequencies $\omega_{0}\pm\omega$
are entangled. These entangled pairs are in the so-called
two-mode squeezed state, and each part alone of an entangled pair
is in a thermal state \cite{Caves-91}. This implies that, when detecting the
squeezing of SPOPOs in a homodyne detection scheme, the local oscillator
should be either a frequency comb with zero frequency shift or a combination
of frequency comb pairs with opposite frequency shifts. 

From (\ref{eq30}), we see that the resonant cavity enhances the squeezing effect
in a frequency range having a width close to the cavity
bandwidth. In practice, to obtain maximum squeezing the
round-trip phase shift $\Delta_{{\rm RT}}$ of the cavity should be
adjusted so that $\Delta_{{\rm RT}}+\Delta_{0}$ approaches a multiple of
$\pi$. As can be seen from (\ref{eq31}), the threshold of SPOPOs corresponds
to the singular point of the input-output transform. Let us assume
$r>0$. Then, we find that the threshold of SPOPO is reached as 
\begin{equation}\label{eq34}
g_{0}=\max(g_{n})\rightarrow{\rm a \cosh}\frac{1+r^{2}}{2r\cos(\Delta_{{\rm RT}}+\Delta_{0})}
\end{equation}
for $\theta=0$ if there exists an integer $n$ such that $|\Delta_{{\rm RT}}+\Delta_{0}-2n\pi|<\pi/2$, or at $\theta=\pi$ when $|\Delta_{{\rm RT}}+\Delta_{0}-(2n+1)\pi|<\pi/2$. This corresponds to the two regimes of the degenerate OPO above threshold \cite{Leindecker-11}.

\noindent \section{Quantum properties of individual pulses}

\noindent Now, let us consider individual pulses. We define the annihilation
operator for the pulse in the \textit{n}th supermode and on the \textit{k}th
period, $\mathscr{T}_{k}$, as: 
\begin{eqnarray}\label{eq35}
\hat{a}_{n,k} & = & \int_{\mathscr{T}_{k}}\psi_{n}^{*}(t-kT_{R})\hat{a}(t){\rm d}t\nonumber \\
 & = & \int_{-\pi}^{+\pi}\frac{e^{ik\theta}}{\sqrt{2\pi}}\hat{a}_{n}(\theta){\rm d\theta},
\end{eqnarray}
whose non-zero commutation relation is: $[\hat{a}_{n,k},\hat{a}_{m,k'}^{\dagger}]\!=\!\delta_{nm}\delta_{kk'}$.
To obtain the second line of (\ref{eq35}), we have used (\ref{eq29}) and (\ref{eq30}). Inversely,
one has 
\begin{equation}\label{eq36}
\hat{a}(t)=\sum_{n,k}\hat{a}_{n,k}\psi_{n}(t-kT_{R}).
\end{equation}
The Hermitian quadrature operators are defined as usual by $\hat{x}_{n,k}=(\hat{a}_{n,k}+\hat{a}_{n,k}^{\dagger})/\sqrt{2}$
and $\hat{p}_{n,k}=i(\hat{a}_{n,k}^{\dagger}-\hat{a}_{n,k})/\sqrt{2},$
with $[\hat{x}_{n,j},\hat{p}_{m,k}]\!=\! i\delta_{nm}\delta_{kk'}$.
From (\ref{eq31}), (\ref{eq32}) and (\ref{eq35}), and assuming $\Delta_{{\rm RT}}+\Delta_{0}=2n\pi$,
we obtain another form of the input-output relation for SPOPOs:
\begin{equation}\label{eq37}
\hat{q}_{n,k}^{(\pm)}=-r\hat{q}_{0,n,k}^{(\pm)}+t^{2}\sum\limits _{s=1}^{\infty}r^{s-1}e^{\pm sg_{n}}\hat{q}_{0,n,k-s}^{(\pm)},
\end{equation}
where $\hat{q}_{n,k}^{(+)}\!=\!\hat{x}_{n,k}$ and $\hat{q}_{n,k}^{(-)}\!=\!\hat{p}_{n,k}$.
If $\Delta_{{\rm RT}}+\Delta_{0}=(2n+1)\pi$, another expression
can be obtained by replacing the reflection coefficient $r$ in the
above expression with $-r$. From (\ref{eq37}), we see that each output pulse
can be considered as the combination of the direct reflection of a first pulse and of the
transmission of the successive intracavity pulses related to pulses that previously
entered the cavity. 

In what follows, we will assume that the input is in a coherent state, including vacuum state. Thus,
the output pulses are in a Gaussian state characterized by their variance
matrix \cite{Dutta-95}. Using (\ref{eq37}), the non-zero matrix elements can be found
to be $\langle\Delta\hat{q}_{n,j}^{(\pm)}\Delta\hat{q}_{n,k}^{(\pm)}\rangle\!=\! V_{n,(j,k)}^{(\pm)}$,
with
\begin{eqnarray}\label{eq38}
V_{n,(j,k)}^{(\pm)} & = & \frac{\delta_{j,k}}{2}\left(r^{2}+\frac{t^{4}e^{\pm2g_{n}}}{1-r^{2}e^{\pm2g_{n}}}\right)\nonumber \\
 & - & \frac{\overline{\delta}_{j,k}}{2}t^{2}\frac{1-e^{\pm2g_{n}}}{1-r^{2}e^{\pm2g_{n}}}(re^{\pm g_{n}})^{|j-k|},
\end{eqnarray}
where $\overline{\delta}_{j,k}=1-\delta_{j,k}$, the first term of
the right hand side is the quadrature variance, and the second is
the covariance. 

Two conclusions can be drawn from this variance
matrix. First, each pulse is in a Gaussian state which is not a pure state: one
can verify that $V_{n,(k,k)}^{(+)}V_{n,(k,k)}^{(-)}\!>\!1/4$ as long
as $g_{n}\!>\!0$. At threshold, the squeezing of the $p-$quadrature
of each pulse is limited, $V_{0,(k,k)}^{(-)}\!=\!2r^{2}/(1+r^{2})$,
whereas $V_{0,(k,k)}^{(+)}\!\rightarrow\!\infty$. Second, there are
pulse-to-pulse quantum correlations and anti-correlations on, respectively,
the $x$- and $p-$quadratures. As the time difference between the pulses increases,
these correlations decrease respectively at a rate of $re^{\pm g_{n}}$.
It should be pointed out, however, that these quantum correlations are not strong enough to ensure bipartite entanglement between any two pulses. This can
be verified on the Duan criterion with the variance matrix given by (\ref{eq38}) \cite{Duan-00}.

Let us finally consider $N$ successive pulses that are in the $n$th
supermode. As discussed above, these pulses are in a mixed Gaussian
state described by the variance matrices of their quadratures $\mathbf{V}_{n}^{(\pm)}(N)$.
It will be useful to look at the minimum eigenvalue problem of the
$p-$quadrature variance matrix: 
\begin{equation}\label{eq39}
\mathbf{V}_{n}^{(-)}(N)\vec{c}_{n,0}^{(N)}=\left[\sigma_{n,0}^{(-)}(N)\right]^{2}\vec{c}_{n,0}^{(N)},
\end{equation}
where $\left[\sigma_{n,0}^{(-)}(N)\right]^{2}$ is the minimum eigenvalue,
and $\vec{c}_{n,0}^{(N)}$ the corresponding eigenvector. With the
explicit expression (\ref{eq38}), they can be found semi-analytically \cite{Gradshteyn-07}.
The normalized elements of $\vec{c}_{n,0}^{(N)}$ read:
\begin{equation}\label{eq40}
c_{n,0}^{(N)}(k)=\mathscr{M}_{n,0}^{(N)}\cos\left[\frac{1}{2}\theta_{n,0}^{(N)}(N\!-\!2k\!-1)\right],
\end{equation}
where $0\!\leq\! k\!<N-1$, $\mathscr{M}{}_{n,0}^{(N)}$ is a normalization
coefficient, and $\theta_{n,0}^{(N)}$ can be found by solving: 
\begin{equation}\label{eq41}
\frac{\cos\theta_{n,0}^{(N)}\left(N+1\right)}{\cos\theta_{n,0}^{(N)}\left(N-1\right)}=re^{-g_{n}},\textrm{ with }0<N\theta_{n,0}^{(N)}<\pi.
\end{equation}
The minimum variance is then: 
\begin{equation}\label{eq42}
[\sigma_{n,0}^{(-)}(N)]^{2}=\frac{1}{2}\left|\frac{r-e^{i\theta_{n,0}^{(N)}}e^{-g_{n}}}{1-re^{i\theta_{n,0}^{(N)}}e^{-g_{n}}}\right|^{2}.
\end{equation}
It is interesting to note that, at large \textsl{$N$} where $\theta_{n,0}^{(\infty)}\rightarrow0$,
it becomes the variance of the \textit{p}-quadrature of the squeezed
frequency comb $\hat{a}_{n}(0)$, and strong squeezing
is re-found as expected.

\noindent \section{Quantum Cramer-Rao bound for time estimation  using SPOPO pulses}

\noindent SPOPOs deliver trains of pulses which can be used as a clock when the pump repetition rate is locked on a time standard, with which  can be measured with a very high accuracy the exact time of occurrence of a given event. It is well-known that quantum noise squeezing or quantum correlations can be used to improve the measurements of physical quantities. The question we address in this section is the determination of the minimum uncertainty that can be reached in the estimation of the temporal positioning of the train of pulses  using the quantum properties of the light produced by SPOPOs, independently of the exact quantity measured on it and on the precise protocol employed for extracting the estimator from the measured quantities : this minimum uncertainty is given by the Quantum Cramer Rao bound \cite{Braunstein-94} that we will now calculate.

Let us consider a probe-field which is time
translated by a small quantity $\tau$: $\hat{a}(t)e^{i\omega_{0}t}\rightarrow\hat{a}(t+\tau)e^{i\omega_{0}(t+\tau)}$,
which can be considered as the result of a unitary transform 
\begin{equation}\label{eq43}
\hat{U}(\tau)=\exp\tau\int_{-\infty}^{+\infty}\hat{a}^{\dagger}(t)(i\omega_{0}+\partial_{t})\hat{a}(t){\rm d}t.
\end{equation}
We want to estimate the time delay, $\tau$, using $N$ successive
SPOPO squeezed pulses on the time interval {[}0, $NT_{0}${]}. Due
to (\ref{eq36}), the reduced system can be shown to evolve under a unitary
transform. In the Schroedinger picture, this reads: 
\begin{equation}\label{eq44}
\hat{\rho}^{(N)}(\tau)=e^{i\hat{\Omega}_{N}\tau}\hat{\rho}^{(N)}(0)e^{-i\hat{\Omega}_{N}\tau},
\end{equation}
where $\hat{\rho}^{(N)}(0)$ is the initial state of the \textsl{$N$}
pulses, described by the variance matrix (\ref{eq38}), and 
\begin{equation}\label{eq45}
\hat{\Omega}_{N}=\sum_{k=0}^{N-1}\sum_{m,n}\Omega_{mn}\hat{a}_{m,k}^{\dagger}\hat{a}_{n,k},
\end{equation}
with $\Omega_{mn}=\int_{\mathscr{T}_{0}}\psi_{m}^{*}(t)(\omega_{0}-i\partial_{t})\psi_{n}(t){\rm d}t.$ 

According to the quantum estimation theory, the estimation
precision of the time delay is limited by the Quantum Cramer-Rao bound, equal to the inverse of the quantum Fisher information $F$ that depends only on the quantum state of the system used in the measurement, given by (\ref{eq44}) in the present case. In Appendix C, we show that, under the assumption that the probe field
is intense and $\tau$ is small, $F$ can be approximately evaluated as: 
\begin{equation}\label{eq46}
F_{N}=\frac{1}{2}\sum_{n}\!(\vec{\alpha}'_{n})^{{\rm T}}\mathbf{\Lambda}_{n}^{(-)}(N)\vec{\alpha}'_{n}.
\end{equation}
where $\mathbf{\Lambda}_{n}^{(-)}(N)\!=\![\mathbf{V}_{n}^{(-)}(N)]^{-1}$,
and the vector elements of $\vec{\alpha}'_{n}$ are defined by: 
\begin{equation}\label{eq47}
\alpha'_{n,k}=\sum_{m}\Omega_{n,m}\langle\hat{a}_{m,k}\rangle,
\end{equation}
which have been assumed to be real. In order to minimize the Quantum Cramer-Rao bound with a fixed mean photon number in the probe-field during the measurement time, the vectors $\vec{\alpha}'_{n}$ should
be proportional to $\delta_{n,0}\vec{c}_{0,0}^{(N)}$, where $\vec{c}_{0,0}^{(N)}$
is defined by (\ref{eq39}). Then, with (\ref{eq47}), we find the optimal probe photon
field envelope, which reads:
\begin{equation}\label{eq48}
\alpha^{(N)}(t)=\alpha_{0}^{(N)}\sum_{k=0}^{N-1}\! c_{0,0}^{(N)}(k)\psi'_{0}(t-kT_{R}),
\end{equation}
where $\psi'_{0}(t)$ is the probe-pulse envelope defined as: 
\begin{equation}\label{eq49}
(\omega_{0}-i\partial_{t})\psi'_{0}(t)=\psi_{0}(t),
\end{equation}
and $\alpha_{0}^{(N)}\!=\!\sqrt{N\overline{n}_{0}(\omega_{0}^{2}+\Delta\omega_{0}^{2})}$,
with $N\overline{n}_{0}=\int|\alpha^{(N)}(t)|^{2}{\rm d}t$ is
the total photon number of mean probe field and $\Delta\omega_{0}^{2}$
its spectral spread \cite{Lamine-08}. Then, the optimal quantum Cramer-Rao
bound reads: 
\begin{equation}\label{eq50}
\Delta\tau^{2}=\frac{[\sigma_{0,0}^{(-)}(N)]{}^{2}}{2N\overline{n}_{0}(\omega_{0}^{2}+\Delta\omega_{0}^{2})}=2\Delta\tau_{{\rm SQL}}^{2}[\sigma_{0,0}^{(-)}(N)]{}^{2},
\end{equation}
where $\Delta\tau_{{\rm SQL}}^{2}$ is the standard quantum limit 
using $N$ coherent pulses, as given in \cite{Lamine-08}. 

Figure 2
\begin{figure}[tb]
 \centerline{ \includegraphics[width=8cm]{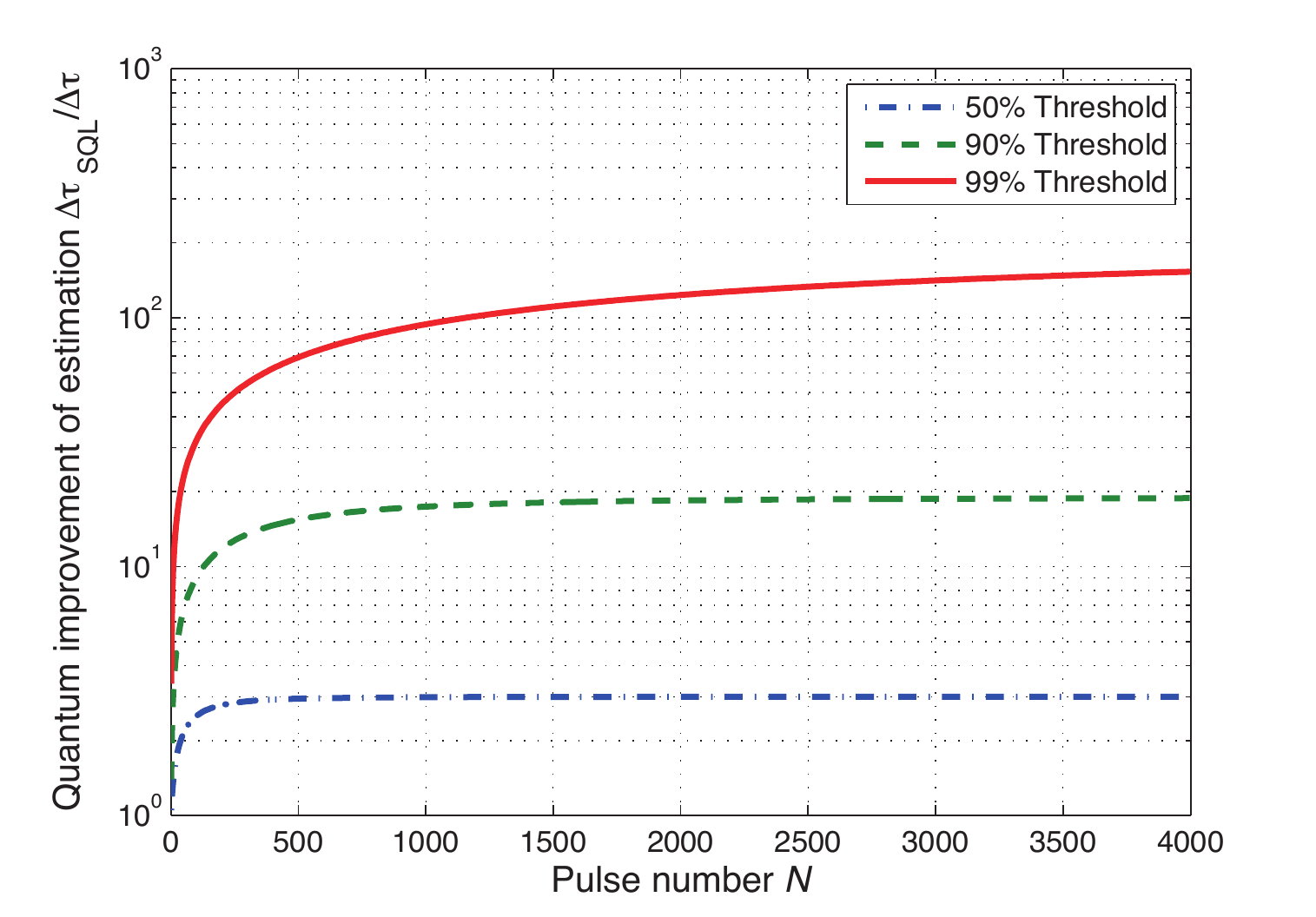}} \caption{Quantum improvement of estimation of a time delay, $\Delta\tau/\Delta\tau_{{\rm SQL}}$,
as a function of pulse number $N$. SPOPO cavity finesse : $\mathcal{F}\approx30$.}
\end{figure}
shows an example of the quantum
improvement of time estimation, $\Delta\tau_{{\rm SQL}}/\Delta\tau$,
 as a function of pulse number $N$ and for different levels of the distance to oscillation threshold. We see that this quantity is always larger than 1, as expected, and that the improvement
increases with the number of pulses used in the measurement until it reaches a limit for large $N$ equal to $\sqrt{2}\sigma_{0,0}^{(-)-1}(\infty)$
that is defined by (\ref{eq42}). We can then know the minimum number of pulses, and therefore the minimum measurement time, needed to reach the limit. In addition, on can verify that
the optimal quantum Cramer-Rao bound (50) is achievable with the
optimal probe field (\ref{eq48}) and the balanced homodyne detection described in ref \cite{Lamine-08},
where the local oscillator is composed of $N$ equal intense coherent
pulses in the supermode $\psi_{0}(t)$.

\noindent \section{Conclusion}

In conclusion, we have presented in this paper a novel time-frequency model for determining the quantum properties of OPOs
below threshold based on a general  input-output formalism that we have applied to the SPOPO case. The impact of the pump CEO and of the cavity round-trip phase on these properties has been studied. We have seen that the frequency combs, or continuous trains of pulses,
form a natural basis for analyzing SPOPOs. As in the pulsed single-pass
configurations, the squeezing of SPOPO is multimode, and the number
of effectively squeezed modes depends on the pump spectrum and the
phase-matching condition. We have shown that SPOPOs generate pairs of entangled frequency combs with frequency shifts symmetric with respect to the central frequencies of PDC.

SPOPOs have also been analyzed in the regime of individual pulses.
We have shown that each individual pulse is in a squeezed thermal
state, and the two quadratures of the pulses are correlated and anti-correlated.
The variance matrices of the quadratures have been analytically evaluated.
Not surprisingly, due to these pulse-to-pulse correlations, strong squeezing is re-found in the multi-pulse regime. Finally,
we have discussed the optimization of quantum improvement of time
estimation with a limited number of SPOPO squeezed pulses. 

In the present paper we have applied the model to the simplest case of a lossless
degenerate type-I ring-cavity in the dispersion compensated configuration, but it can serve as a framework for analyzing more complex configurations which are closer to real experimental conditions, such as higher-order intra-cavity dispersion effects or the influence of the transverse spatial dependence.

\section*{Acknowledgments}
We acknowledge the financial support of the Future and Emerging Technologies (FET) programme within the Seventh Framework Programme for Research of the European Commission, under the FET-Open grant agreement HIDEAS, number FP7-ICT-221906; of the ANR project QUALITIME and of the ERC starting grant FRECQUAM.

\noindent \section*{Appendix A}

\noindent Here, we first show how to find the common eigenvector functions
of $\boldsymbol{\mathcal{U}}_{0}={\rm Diag}(\mathcal{U}_{0},\mathcal{U}_{0}^{*})$
defined in (\ref{eq25}) and
\begin{equation}
\boldsymbol{\mathcal{H}}_{0}=\ln\boldsymbol{\mathcal{G}}_{0}=\left(\begin{array}{cc}
0 & \mathcal{S}_{0}\\
\mathcal{S}_{0}^{\mathtt{*}} & 0
\end{array}\right),
\end{equation}
that is defined by (\ref{eq15}). Since $\boldsymbol{\mathcal{U}}_{0}$ is
unitary, its eigenvalues are complex and on the unit circle, i.e., 
\begin{equation}
\boldsymbol{\mathcal{U}}_{0}\boldsymbol{f}(\theta)=e^{i\theta}\boldsymbol{f}(\theta).
\end{equation}
Writing $\boldsymbol{f}(\theta)=(f_{1}(\theta),f_{2}^{*}(\theta))^{{\rm T}}$,
from (\ref{eq25}), we have
\begin{eqnarray}
f_{1}(\theta,t+T_{0}) & = & f_{1}(\theta,t)e^{i(\Delta_{0}+\theta)},\nonumber \\
f_{2}(\theta,t+T_{0}) & = & f_{2}(\theta,t)e^{i(\Delta_{0}-\theta)},
\end{eqnarray}
which means they are quasi $T_{0}-$periodic functions. Let us assume
$\boldsymbol{f}(\theta)$ is also the eigenvector of $\boldsymbol{\mathcal{H}}_{0}$,
i.e., $\boldsymbol{\mathcal{H}}_{0}\boldsymbol{f}(\theta)=g(\theta)\boldsymbol{f}(\theta)$,
where the eigenvalue $g(\theta)$ is a real number because $\boldsymbol{\mathcal{H}}_{0}$
is Hermitian. From $\boldsymbol{\sigma}_{1}\boldsymbol{\mathcal{H}}_{0}\boldsymbol{\sigma}_{1}=-\boldsymbol{\mathcal{H}}_{0}$,
we see that $\boldsymbol{\sigma}_{1}\boldsymbol{f}(\theta)$ is also
the eigenvector of $\boldsymbol{\mathcal{H}}_{0}$ with eigenvalue
$-g(\theta)$. Moreover, it is easy to see that $\boldsymbol{\sigma}_{1}\boldsymbol{f}(\theta)$
is also the eigenvector of $\boldsymbol{\mathcal{U}}_{0}$. Thus,
both $\boldsymbol{f}^{(+)}(\theta)=\boldsymbol{f}(\theta)$ and $\boldsymbol{f}^{(-)}(\theta)=\boldsymbol{\sigma}_{1}\boldsymbol{f}(\theta)$
are common eigenvectors of $\boldsymbol{\mathcal{U}}_{0}$ and $\boldsymbol{\mathcal{H}}_{0}$:
\begin{eqnarray}
\boldsymbol{\mathcal{H}}_{{\rm 0}}\boldsymbol{f}^{(\pm)}(\theta) & = & \pm g(\theta)\boldsymbol{f}^{(\pm)}(\theta),\\
\boldsymbol{\mathcal{U}}_{{\rm 0}}\boldsymbol{f}^{(\pm)}(\theta) & = & e^{i\theta}\boldsymbol{f}^{(\pm)}(\theta).
\end{eqnarray}
Now, since $\boldsymbol{\mathcal{H}}_{0}^{2}\boldsymbol{f}^{(\pm)}(\theta)\!=\! g^{2}(\theta)\boldsymbol{f}^{(\pm)}(\theta)$,
with (53), we find:
\begin{equation}
\mathcal{S}_{0}\mathcal{S}_{0}^{*}f_{1/2}(\theta)=g^{2}(\theta)f_{1/2}(\theta).
\end{equation}
From (\ref{eq22}) and (\ref{eq24}), this means
\begin{equation}
\int_{\mathscr{T}_{0}}K_{2}(t,t')f_{1/2}(\theta,t'){\rm d}t'=g^{2}f_{1/2}(\theta,t),
\end{equation}
with $\mathcal{K}_{2}=\mathcal{S}_{C}\mathcal{S}_{C}^{*}$ and $t\in\mathscr{T}_{0}$,
where we have explicitly eliminated the $\theta-$dependence of $g$.
Since $\mathcal{K}_{2}$ is Hermitian $L^{2}$ \cite{Parker-00}, its eigenfunctions,
\{$\psi_{n}(t)$\} defined by: 
\begin{equation}
\int_{0}^{T_{0}}K_{2}(t,t')\psi_{n}(t'){\rm d}t'=g_{n}^{2}\psi_{n}(t),
\end{equation}
form a complete orthonormal set. Thus, with (55), (58) and
(\ref{eq24}), we find (\ref{eq29}). 

\noindent \section*{Appendix B}

\noindent It is easy to see that the functions of the set \{$f_{n}(\theta,t)$\}
are orthonormal. To show the completeness, one first notes that for
an arbitrary well-behaved function, say $\alpha(t)$, its inverse
Fourier transform can be written as: 
\begin{eqnarray*}
\alpha(t) & = & \frac{1}{2\pi}\int_{-\infty}^{+\infty}\underline{\alpha}(\omega)e^{i\omega t}{\rm d}\omega\\
 & = & \frac{1}{2\pi}\sum_{k}\int_{(k-1/2)\Omega_{0}+\Delta_{0}/T_{0}}^{(k+1/2)\Omega_{0}+\Delta_{0}/T_{0}}\underline{\alpha}(\omega)e^{i\omega t}{\rm d}\omega\\
 & = & \int_{-\Omega_{0}/2}^{+\Omega_{0}/2}\alpha_{B}(\omega,t)e^{i(\omega+\Delta_{0}/T_{0})t}{\rm d}\omega,
\end{eqnarray*}
where $\Omega_{0}\!=\!2\pi T_{0}^{-1}$, and 
\[
\alpha_{B}(\omega,t)=\frac{1}{2\pi}\sum_{k}e^{ik\Omega_{0}t}\underline{\alpha}(k\Omega_{0}+\omega+\Delta_{0}/T_{0})
\]
is the so-called Bloch amplitude, which is $T_{0}-$periodic. On the
other hand, for each given $\theta$, the set of functions \{$g_{n}(\theta,t)$\},
defined as
\begin{eqnarray*}
g_{n}(\theta,t) & = & \sqrt{2\pi}f_{n}(\theta,t)e{}^{-iT_{0}^{-1}(\theta+\Delta_{0})t}\\
 & = & \sum\limits _{k=-\infty}^{+\infty}\psi_{n}(t-kT_{0})e^{-iT_{0}^{-1}(t-kT_{0})(\theta+\Delta_{0})},
\end{eqnarray*}
is $T_{0}-$periodic and is complete orthonormal basis for all $T_{0}-$periodic
functions due to the completeness of \{$\psi_{n}$\}. Thus, we can
project $\alpha_{B}(\omega,t)$ on the basis \{$g_{n}(\omega T_{0},t)$\}:
\[
\alpha_{B}(\omega,t)=\sum_{n}\alpha'_{n}(\omega T_{0})g_{n}(\omega T_{0},t),
\]
with $\alpha'_{n}(\omega T_{0})\!=\!\int_{\mathscr{T}_{0}}g_{n}^{*}(\omega T_{0},t)\alpha_{B}(\omega,t){\rm d}t$.
Therefore,
\begin{eqnarray*}
\alpha(t) & = & \int_{-\Omega_{0}/2}^{+\Omega_{0}/2}\sum_{n}\alpha'_{n}(\omega T_{0})g_{n}(\omega T_{0},t)e^{i(\omega+\Delta_{0}/T_{0})t}{\rm d}\omega\\
 & = & T_{0}^{-1}\int_{-\pi}^{+\pi}\sum_{n}\alpha'_{n}(\theta)g_{n}(\theta,t)e^{iT_{0}^{-1}(\theta+\Delta_{0})t}{\rm d}\theta\\
 & = & \frac{\sqrt{2\pi}}{T_{0}}\int_{-\pi}^{+\pi}\sum_{n}\alpha'_{n}(\theta)f_{n}(\theta,t){\rm d}\theta\\
 & = & \int_{-\pi}^{+\pi}\sum_{n}\alpha_{n}(\theta)f_{n}(\theta,t){\rm d}\theta,
\end{eqnarray*}
where we have made the substitution $\theta\rightarrow\omega T_{0}$
in the second line, and
\begin{eqnarray*}
\alpha_{n}(\theta) & = & \frac{\sqrt{2\pi}}{T_{0}}\alpha'_{n}(\theta)=\int_{-\infty}^{+\infty}f_{n}^{*}(\theta,t)\alpha(t){\rm d}t.
\end{eqnarray*}
So, we have shown that any well-behaved function, for which a Bloch amplitude
exists, can be written as the combination of the functions of \{$f_{n}(\theta,t)$\}.
Thus, frequency combs indeed form a complete orthonormal set.

\noindent \section*{Appendix C}

\noindent Since our system, evolving as (\ref{eq44}), is in a mixed state,
the Quantum Fisher Information should be evaluated as \cite{Braunstein-94}: 
\begin{equation}
F_{N}={\rm tr}\hat{\rho}^{(N)}(0)\hat{L}_{N}^{2},
\end{equation}
where $\hat{L}_{N}$ is the symmetric logarithmic derivative that
verifies: 
\begin{equation}
\frac{1}{2}[\hat{L}_{N}\hat{\rho}^{(N)}(0)+\hat{\rho}^{(N)}\hat{L}_{N}]=i[\hat{\Omega}_{N},\hat{\rho}_{N}(0)].
\end{equation}
We can replace $\hat{\Omega}_{N}$ by $\Delta\hat{\Omega}_{N}=\hat{\Omega}_{N}-\langle\hat{\Omega}_{N}\rangle$
in the above commutator. Assuming the field is intense and $\tau$
is small, from (\ref{eq45}), we have approximately:
\begin{equation}
\Delta\hat{\Omega}_{N}=\sqrt{2}\sum_{k=0}^{N-1}\sum_{n}\alpha'_{n,k}\Delta\hat{x}_{n,k},
\end{equation}
where $\alpha'_{n,k}=\sum_{m}\Omega_{n,m}\langle\hat{a}_{m,k}\rangle$
has been assumed to be real. 

Since the sub-systems of supermodes are uncorrelated, $\hat{\rho}^{(N)}(0)$
can be written as a direct product of sub-systems: $\hat{\rho}^{(N)}(0)=\prod_{n=0}^{\infty}\otimes\hat{\rho}_{n}^{(N)}$,
where $\hat{\rho}_{n}^{(N)}$ is the state of the $n$th supermode.
It is convenient to project $\hat{\rho}_{n}^{(N)}$ on the eigenstates
of the $p-$quadratures \cite{Dutta-95}:
\begin{equation}
\hat{\rho}_{n}^{(N)}=\int_{\Re^{2N}}|\vec{p}\rangle\tilde{W}_{n}(\vec{p},\vec{p}')\langle\vec{p}'|{\rm d}^{N}\vec{p}{\rm d}^{N}\vec{p}',
\end{equation}
with
\begin{equation}
\tilde{W}_{n}(\vec{p},\vec{p}')=\int_{\Re^{N}}{\rm d}^{N}\vec{x}W_{n}(\vec{x},\frac{\vec{p}+\vec{p}'}{2})e^{-i\vec{x}\cdot(\vec{p}-\vec{p}')},
\end{equation}
where $W_{n}(\vec{x},\vec{p})$ is the Wigner function. For our case,
it is a Gaussian function of $2N$ variables.
\begin{equation}
W_{n}(\vec{x},\vec{p})=\frac{e{}^{-\frac{1}{2}(\Delta\vec{x}_{n}^{{\rm T}}\mathbf{\Lambda}_{n}^{(+)}(N)\Delta\vec{x}_{n}+\Delta\vec{p}_{n}^{{\rm T}}\mathbf{\Lambda}_{n}^{(-)}(N)\Delta\vec{p}_{n})}}{|2\pi\mathbf{V}_{n}^{(+)}(N)|^{\frac{1}{2}}|2\pi\mathbf{V}_{n}^{(-)}(N)|^{\frac{1}{2}}},
\end{equation}
where \textit{$\mathbf{\Lambda}_{n}^{(\pm)}(N)\!=\![\mathbf{V}_{n}^{(\pm)}(N)]^{-1}$},
and $\Delta\vec{x}_{n}=\vec{x}-\vec{x}_{n}$ and $\Delta\vec{p}_{n}=\vec{p}-\vec{p}_{n}$,
with $\vec{x}_{n}$ and $\vec{p}_{n}$ being the mean values. Now,
since $\langle p'|\hat{x}|p\rangle=-i\partial_{p}\delta(p-p')$, we
have from (62):
\begin{equation}
\langle\vec{p}|[\Delta\hat{x}_{n,k},\hat{\rho}_{n}^{(N)}]|\vec{p}'\rangle=i(\partial_{p_{k}}+\partial_{p'_{k}})\tilde{W}_{p}(\vec{p},\vec{p}').
\end{equation}
Then, it is straightforward to show that the symmetric logarithmic
derivative reads: 
\begin{equation}
\hat{L}_{N}=\frac{1}{\sqrt{2}}\sum_{n}\sum_{j,k=0}^{N-1}\Lambda_{n,(k,j)}^{(-)}(N)\alpha'_{n,k}\Delta\hat{p}_{n,j}.
\end{equation}
 Hence, the Quantum Fisher Information (57) now reads:
\begin{equation}
F_{N}=\frac{1}{2}\sum_{n}\sum_{j,k=0}^{N-1}\Lambda_{n,(k,j)}^{(-)}(N)\alpha'_{n,k}\alpha'_{n,j}.
\end{equation}

\end{document}